\documentclass[10pt,aps,twocolumn, superscriptaddress]{revtex4-1}

\usepackage{amsmath,amsthm}
\usepackage{amssymb}
\usepackage{amscd}
\usepackage{wasysym}
\usepackage[a]{esvect}
\usepackage[normalem]{ulem}
\usepackage[ansinew]{inputenc}
\usepackage[T1]{fontenc}
\usepackage{ae,aecompl}
\usepackage{hyperref}
\usepackage{enumitem}
\usepackage[ruled,vlined]{algorithm2e}
\usepackage{bbold}
\usepackage[pdftex]{graphicx}
\graphicspath{{pics/}}

\usepackage{hyperref}
\hypersetup{
    colorlinks=true,       
    linkcolor=red,          
    citecolor=green,        
    filecolor=magenta,      
    urlcolor=cyan           
}

\usepackage{url}

\usepackage[textsize=tiny]{todonotes}
\usepackage{listings}
\usepackage{color}
\definecolor{red}{rgb}{1,0,0}
\definecolor{blue}{rgb}{0,0,1}
\definecolor{green}{rgb}{0,0.5,0}
\definecolor{magenta}{HTML}{90004F}

\renewcommand{\vec}[1]{\vv{#1}}
\newcommand{\fig}{FIG~}

\newcommand{\mb}[1]{\mathbb{#1}}
\newcommand{\V}{\mb{V}}
\newcommand{\E}{\mb{E}}
\newcommand{\R}{\mb{R}}
\newcommand{\N}{\mathcal{N}}
\newcommand{\flownet}{\mathcal{F}}

\newcommand{\doubimply}{\Leftrightarrow}

\newcommand{\bigO}[1]{\mathcal{O}\left(#1\right)}
\renewcommand{\tilde}{\widetilde}

\newcommand{\metaGraph}[1]{\tilde{#1}}

\newcommand{\linv}{L^+}

\newtheorem{lemma}{Lemma}

\newtheorem{claim}{Claim}

\newtheorem{conjecture2}{Heuristic}

\newtheorem{definition}{Definition}

\begin{document}
\bibliographystyle{apsrev}
\title{Predicting Braess' Paradox in Supply and Transport Networks}

\author{Debsankha Manik}
\affiliation{Chair for Network Dynamics, Institute for Theoretical Physics
and Center for Advancing Electronics Dresden (cfaed),
Technical University of Dresden, 01062 Dresden, Germany}

\author{Dirk Witthaut}
\affiliation{Forschungszentrum J\"ulich, Institute for Energy and Climate Research -
    Systems Analysis and Technology Evaluation (IEK-STE),  52428 J\"ulich, Germany}
\affiliation{Institute for Theoretical Physics, University of Cologne,
    50937 K\"oln, Germany}

\author{Marc Timme}
\affiliation{Network Dynamics, Max Planck Institute for Dynamics and Self-Organization (MPIDS), 37077 G\"ottingen, Germany}
\affiliation{Chair for Network Dynamics, Institute for Theoretical Physics
and Center for Advancing Electronics Dresden (cfaed),
Technical University of Dresden, 01062 Dresden, Germany}

\date{\today}

\begin{abstract}
 Reliable functioning of supply and transport networks fundamentally support
 many non-equilibrium dynamical systems, from biological organisms and
 ecosystems to human-made water, gas, heat, electricity and traffic networks.
 Strengthening an edge of such a network lowers its resistance opposing a flow
 and intuitively improves the robustness of the system's function. If, in
 contrast, it deteriorate operation by overloading other edges, the
 counterintuitive phenomenon of  \emph{Braess' paradox} emerges. How to predict
 which edges enhancements may trigger Braess' paradox remains unknown to date.
 To approximately locate and intuitively understand such Braessian edges, we
 here present a differential perspective on how enhancing any edge impacts
 network-wide flow  patterns. First, we exactly map the prediction problem to a
 dual problem of electrostatic dipole currents on networks such that
 simultaneously finding  \textit{all} Braessian edges is equivalent to finding
 the currents in the resistor network resulting from a constant current across
 one edge. Second, we propose a simple approximate criterion -- rerouting
 alignment -- to efficiently predict Braessian edges, thereby providing an
 intuitive topological understanding of the phenomenon. Finally, we show how to
 intentionally weaken Braessian edges to mitigate network overload,
 with beneficial consequences for network functionality.
\end{abstract}

\maketitle

\section{Introduction}
Properly functioning supply networks essentially underlie our everyday lives. They enable the transport of nutrients and fluids \cite{Kati10} in biological organisms such as our body,
the supply of electric energy in power grids \cite{kundur1994power}, the
transport of people and goods in traffic networks
\cite{nagurney2000sustainable} as well as the flow of information through
communication networks such as the internet. Structural changes of supply networks impact their core functionality.
Increasing the strength of an edge or adding a new edge to a supply network constitutes a common strategy for improving its overall transport performance and to adapt it to current or future needs \cite{amin2008challenges}. However, not every enhanced edge actually improves performance.

Indeed, already in 1968, Braess \cite{Brae68} highlighted an intriguing phenomenon in traffic networks: that opening a new street can worsen overall traffic as each individual tries to selfishly optimize their own travel time.
Thus, adding certain edges may decrease the transport performance of supply and transport networks, a collective phenomenon today known as Braess' paradox.

In its extreme form, Braess' paradox may induce a complete loss of operating state of the supply network and thus a total collapse of its functionality, see e.g.~\cite{witthaut_nonlocal_2013}. Weaker forms of Braess' paradox reduce system performance and causes higher stress in the network, e.g. reduce overall traffic flow in a street network \cite{Brae05} or reduce system stability \cite{colombo_braess_2016}. Since its first identification, Braess' paradox has been shown to prevail across many networked systems, including traffic networks, DC electrical circuits, AC electricity grids and other oscillatory networks, linear supply networks,  discrete message passing systems, and two-dimensional electron gases
\cite{cohen_paradoxical_1991,12braess, witthaut_nonlocal_2013,
nagurney_physical_2016, toussaint_counterintuitive_2016}. Notable theoretical results achieved over the
decades include necessary and sufficient
conditions for the occurrence of Braess' Paradox \cite{frank_braess_1981, steinberg_prevalence_1983,
dafermos_traffic_1984, korilis_avoiding_1999,
azouzi_properties_2002, nagurney2010negation}, its prevalence even if
individual agents behave non-selfishly \cite{pas_braess_1997}, and the
likelihood of occurrence in large random networks \cite{valiant_braesss_2010}.
An algorithmic heuristic of identifying Braessian edges in traffic networks have been proposed by \cite{Bagloee:2014BraessPredictingHeuristics}.
Coletta and Jacquod
\cite{coletta_linear_2016} recently showed how to predict which edges, if enhanced, cause Braess' paradox, i.e. which edges are ``Braessian'' for heterogeneous one-dimensional chain topologies
Yet, a general theory to better understand and to predict which individual edges are Braessian in a network is still missing to date.

To intuitively understand where and why Braess' paradox occurs,  we here take a new direction towards such a theory of predicting Braessian edges in networks with a wide class of dynamics. We propose an alternative perspective and consider Braess' paradox in terms of increasing maximal flows and inducing potential overloads of edges due to \emph{differential} changes of the network structure. Accordingly, we call an edge \textit{Braessian} if
infinitesimally increasing its strength yields an increase in the maximum flow in the network. 

The problem of identifying such differentially Braessian 
edges maps exactly to a dual problem from electrostatic theory: that of identifying the direction of network currents induced by one dipole current on the maximum-flow carrying
edge. Guided by the intuition resulting from this mapping, we propose an intuitive approximate graph theoretic \emph{predictor} of Braessian edges based on the direction of rerouted flows if the maximum flow carrying edge is removed.  
We illustrate the inverse consequence of these results to indicate ways to intentially reduce the strength of a
Braessian edge to recover an operating state of originally overloaded networks.
The insights thus not only further our theoretical understanding of Braess' paradox  by providing intuitive insights about where to expect Braessian edges and provide drastic computational simplifications, they also offer practical advice on how to keep supply and transport networks functional.

\section{Guiding Background}
For the theory we develop below, we consider a broad class of supply and
transport networks as occurring in natural and engineered systems.  Before we
provide more details, let us first mention some key properties of the networks
constituting that class.
By \textit{supply networks} we refer to graphs $G(\V,\E)$ of vertices (nodes) $i\in\{1,\ldots,N\}=:\V$ and edges (lines) $(i,j)\in \V\times \V$ having
the following additional vertex and edge properties:
\begin{enumerate}
\item \textit{flows} $F_{ij}\in\R$ across an edge $(i,j)$ quantify the amount of material or energy transported across that edge per unit time;
\item \textit{scalar vertex variables} $\varphi:\V\rightarrow\R,\ i\mapsto \varphi_i$ define potential functions in the sense of physical potentials such that these scalars $\varphi_i$ constitute state variables making the resulting flows conservative
and
\item \textit{edge strengths} $K_{ij}\in\R$ are the inverse of edge resistances that oppose flows, edge strengths are thus generalized susceptances known from DC electric networks. 
\end{enumerate}
These quantities are related by 
\begin{equation}
    F_{ij}=K_{ij}f(\varphi_j-\varphi_i)
\end{equation}
where $f:\mathbb{R}\rightarrow \mathbb{R}$ is a differentiable, strictly monotonic and odd function of the (potential) difference $(\varphi_j-\varphi_i)$ of the state variables at the vertices $i$ and $j$ the edge is incident to.

\subsection{Linear supply networks}
For linear $f$ and thus $f(x)=x$, without loss of generality, we obtain the most basic model setting where the flows from $j$ to $i$ are given by
\begin{align}
    \label{def:linsuppnet}
    F_{ij}&=K_{ij}(\varphi_j-\varphi_i).
\end{align}
Such networks provide suitable approximate models for electric circuits, water,
gas and heat supply networks as well as biological systems such as plant
venation networks supplying, e.g. plant leaves \cite{stott2009dc,Kati10}.

For electric circuits, \eqref{def:linsuppnet} represents Ohm's law,
approximating the current $F_{ij}$ flowing through a conductor for a given
potential difference $\varphi_j-\varphi_i$ between its end nodes by a linear
function with the current proportional to the conductance, i.e. the inverse
resistance, $K_{ij}:=1/R_{ij}$. In reality, the conductance depends on several
factors, including the temperature of the conductor which itself depends on the
current flowing through it. Thus, a linear relation approximates the actual
nonlinear relation between voltage (potential difference) and current (flow) at
a given operating point.

\subsection{Nonlinear supply networks}
Any nonlinearity of $f$ characterizes system-specific details beyond the
linearization of a network near a given operating point. For instance, coupled
swing equations exhibit sinusoidal $f$ and model lossless electric AC
transmission grids. In that model class, each power generator and each consumer
is  modeled as a synchronous machine, and thus assigned a phase $\theta_j$, a
moment of itertia $M_j$, a damping constant $D_j$ as well as the power produced
(for generators) or consumed (for consumers) $P_j$ \cite{witthaut2022RMP,Fila08}.

The equation of motion at each node is given by
\begin{align}
    \label{def:swingeq}
    M_j\frac{d^2\theta_j}{dt^2} + D_j \frac{d\theta_j}{dt} =
     P_j        +\sum_{k=1}^N  K_{jk} \sin(\theta_k-\theta_j),
\end{align}
where the edge strengths $K_{ij}$ depend on the grid voltage approximated to be
constant in time and the admittance of the transmission line.

During the steady operation of the power grid, the flow of electrical power from node $j$ to node $i$ is given by
\begin{align}
    \label{def:swingflow}
    F_{ij}&=K_{ij}\sin\left(\varphi_j-\varphi_i\right),
\end{align}
identifying $\theta_i=\varphi_i$.

Model flows with nonlinear $f$ equally represent flows that can be assigned to
dynamical systems that originally do not model real supply or transport
networks.  For instance, equations \eqref{def:swingflow} define abstract flows
for the Kuramoto model \cite{Kura84,Aceb05} with variables $\varphi_i(t)$ satisfying
\begin{equation}
 \frac{d\varphi_i}{dt} =
      \omega_i+\sum_{j=1}^N K_{ij} \sin(\varphi_j-\varphi_i),
\end{equation}
where $\omega_j$ are the natural frequencies of each node.
The Kuramoto model constitutes a paradigmatic model of weakly coupled, strongly
attracting limit cycle oscillators and does not include any types of material,
energy or other flows.  We may thus \textit{assign} flows to systems that are
not models of supply or transport networks to uncover system properties
employing approaches for supply networks, for instance the approximate
prediction scheme for Braessian edges presented below.

\section{Braess' paradox in supply networks}
\label{sec:definitions}
Now, equipped with basic ideas about the system class considered, let us define \emph{conservative supply networks} and introduce an \emph{infinitesimal}
perspective onto Braess' paradox, on which we base the core results of this article.

\paragraph{Conservative supply networks.}
As sketched above, supply networks are graphs
whose edges model the transport of a certain quantity -- which can be matter, energy or information. This quantity enters the system
through a subset of \emph{source} nodes and  exits it through
another subset of \emph{sink} nodes. We formalize this in the
following definition.
\begin{definition}[Supply network] \label{def:supplynet}
    Let $G(\V,\E)$ be a graph with the vertex set $\V=(v_1,v_2,\cdots,v_N)$
    and the edge set $\E=(e_1,e_2,\cdots,e_M)$ and let us denote the input (current) at each node $j \in \V$ as $I_j$ and the flow \emph{from $j$ to $i$}
    across each edge $e=(i,j)\in\E$ as $F_e \equiv F_{ij}$.
    Moreover, let $\vec{I}:=(I_1, I_2,\cdots,I_N)$ and $\vec{F}=(F_{e_1},F_{e_2},\cdots,F_{e_M})$.
    Then the tuple $(G,\vec{I},\vec{F})$ is called a supply network.
\end{definition}
Let us focus one supply networks where the flow is
conserved such that the continuity equation
\begin{equation}
    \label{eq:flowcon}
    I_j = \sum_{(i,j)\in\E} F_{ij}, \text{ for all } j \in\V
\end{equation}
holds. It means that the input at each node equals the total outward flow through
all the edges that node is part of. We remark that both, the $I_j$ and the $F_{ij}$ may be positive, negative or zero.

\begin{definition}[Conservative supply network]\label{def:conflownet}
    A supply network $(G, \vec{I}, \vec{F}^{\textsf{con}})$ is called a
    \emph{conservative} supply network if the continuity equation \eqref{eq:flowcon} is satisfied and the flow $F_{ij}$ across any edge $(i,j)$ is a
    \emph{monotonically increasing, continuous, differentiable and odd} function of
    the \emph{difference between a certain vertex property} across the edge
    $(i,j)$,
    \begin{alignat}{2}
        F_{ij} &=&& K_{ij} f(\varphi_j - \varphi_i) \label{eq:conserv-prop-flow}\\
           f(y)&>&& f(x) \doubimply y > x\label{eq:f-monotonic-cond} \\
           f(-x)& =&& -f(x) \\
           K_{ij}&=&&K_{ji}
    \end{alignat}
    for all $i,j\in\V$ and all $x,y$ in the domain of $f$.
\end{definition}
We note that the oddness of $f$ together with the symmetry of the $K_{ij}$ implies the flow directionality condition $F_{ij}=-F_{ji}$ that makes the flows well-defined.

\paragraph{Differential Braess' paradox.}
Most works on Braess' paradox, including the first work by Braess \cite{Brae68}
define it as a some form of ``decrease in performance'' of a supply network upon \emph{adding
an edge}. We here broaden this perspective by considering the
\emph{infinitesimal strengthening} of an edge, a continuous procedure, instead of adding an edge, a discrete procedure.
\begin{figure}[!htp]
    \begin{center}
        \includegraphics[width=\columnwidth]{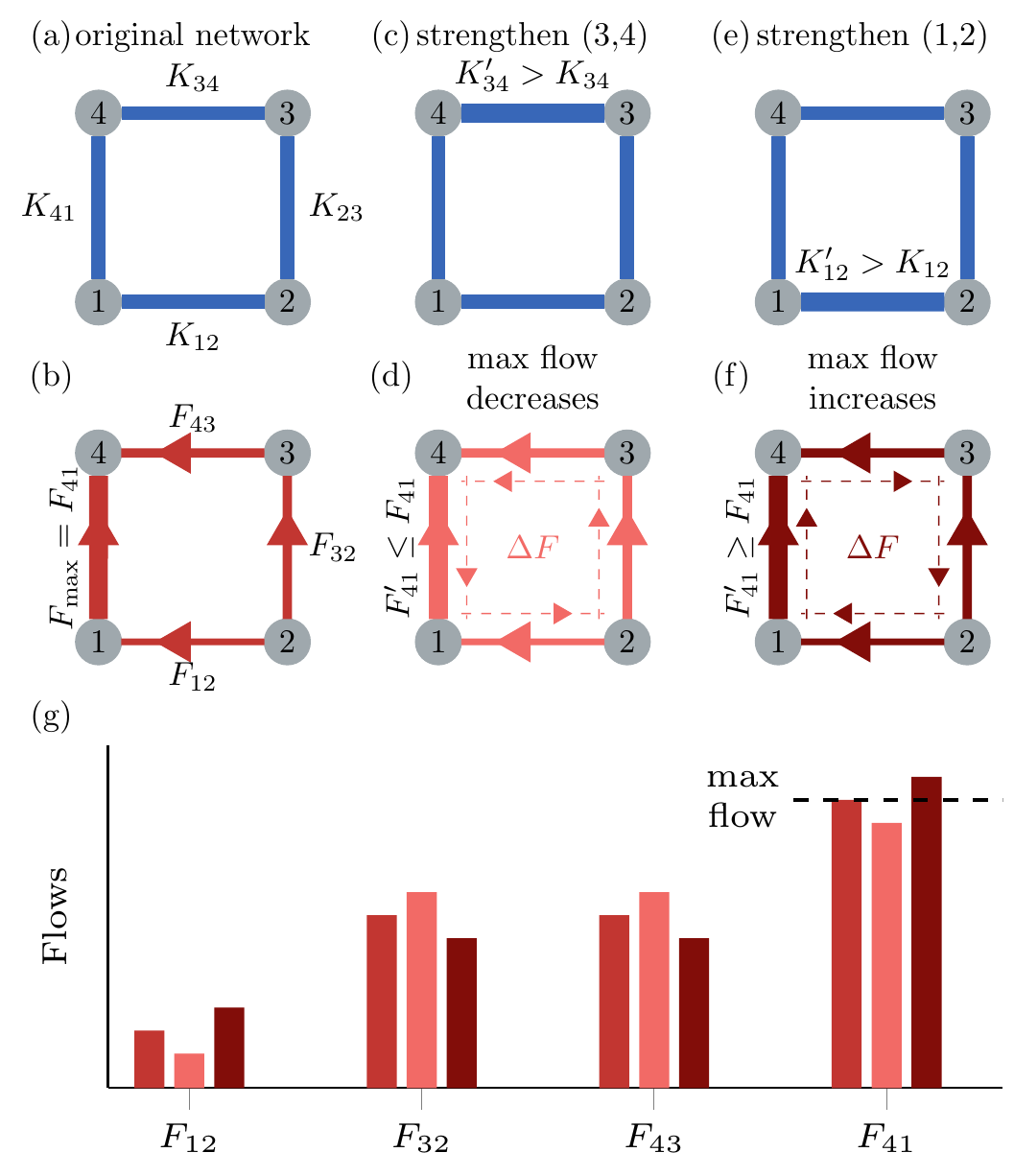}
        \caption{\textbf{Braessian edge that aligns with differential flow change.} (a) Simple  supply network with four nodes (gray disks) with heterogeneous inputs $I_i$ (not illustrated), four edges (blue), and identical edge strengths $K_{ij}=K_0$. (b) Maximum flow is
        across the edge at the left. (c) Top edge slightly
        strengthened. (d) In response, the incremental flow changes along a \emph{cycle},
        (dashed red lines) that disaligns with and thus decreases maximum flow. Hence, the top
        edge is not Braessian. (e) Bottom edge is slightly strengthened instead and (f)  maximum flow increases. Hence, the bottom edge is Braessian. (g) Bar chart showing, for each edge, the original flows
        (red), flows due to strengthening top edge (light red), and due to
        strengthening bottom edge (dark red).} \label{fig:be-def}
    \end{center}
\end{figure}
\begin{definition}[Braessian edge]
    \label{def:be}
    In a supply network $\flownet=(G, \vec{I}, \vec{F})$, let the maximum flow
    be across the edge $(s,t)$, $\max_{(i,j)\in \E} |F_{ij}|=|F_{st}|$.
    After increasing the strength of only one edge $(a,b)$ by a small amount,
    $K'_{ab}=K_{ab}+\kappa$, let the new flows across the edges be
    $F'_{ij}=F_{ij}+\delta F_{ij}$.  The edge $(a,b)$ is called Braessian if
    and only if
    \begin{align}
        \label{eq:def-be}
        |F'_{st}| > |F_{st}| 
    \end{align}
    as $\kappa\rightarrow 0.$
\end{definition}
We note that the condition \eqref{eq:def-be} is equivalent to $ F_{st} \delta
F_{st} > 0$. We illustrate this definition in \fig \ref{fig:be-def} by means of
a simple four-node supply network. The maximum flow is across the left edge
$(1,4)$. Upon increasing the strength of the top edge infinitesimally (panel
b), the resulting incremental flow change at the maximum flow edge is
\emph{anti-aligned}) with (i.e. not in the same direction as) its original
flow. Thus the maximum flow increases upon increasing the strength of the top
edge, which makes it non-Braessian. However, the top edge is Braessian, because
upon increasing its strength (panel c), the incremental flow change at the left
edge is \emph{aligned} to the original (maximum) flow.

\section{Electrostatic analog and key symmetry}
\label{sec:electro-analog}
Intriguingly, the problem of determining Braessian edges has a simple
electrostatic analog that is crucial for our core result of understanding
Braess' Paradox (BP) based on the network topology.  If a single edge of a
conservative supply network is strengthened, the resulting flow changes are
equal to the currents in a specifically constructed resistor network, where a
constant current source is placed across the edge with the maximum flow. We
will now present this equivalence in detail.

\begin{figure}[!htp]
    \begin{center}
        \includegraphics[width=\columnwidth]{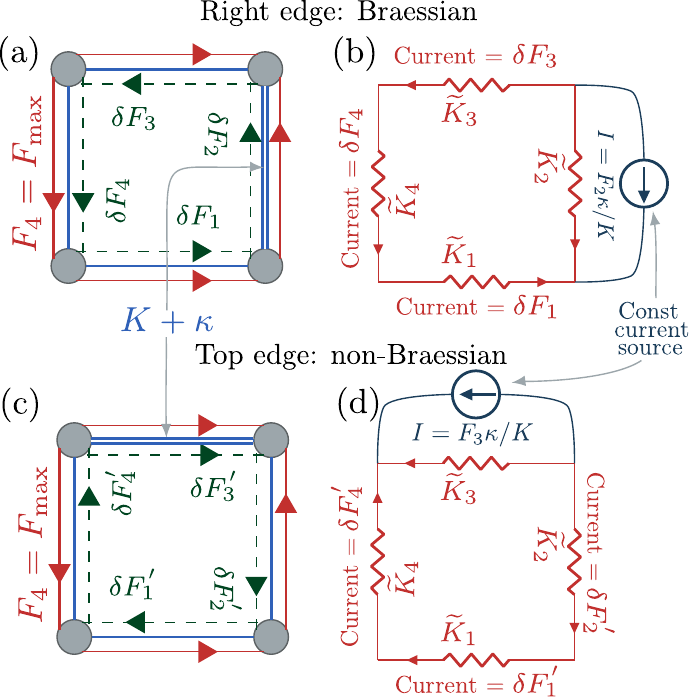}
        \caption{Duality between flow changes and currents in a resistor
        network utilized to identify Braessian edges.  (a) Flows in a four-node
        network (solid red edges) and incremental flow changes (dashed green
        edges) on strengthening the edge on the right. (b) A resistor network
        with resistances given by \eqref{eq:red-capacity} and a constant
        current source connected across the right edge. The resulting currents
        in (b) equal the incremental flow changes in (a) at all edges except the
        right edge. The current at the left edge is aligned to the 
        maximum flow, hence Braess' paradox occurs.  (c) -(d) The same as (a)-(b), but with the top edge strengthened. The current at the left edge is anti-aligned to the 
        maximum flow, hence there is no Braess' paradox.}
        \label{fig:resistor-equiv}
    \end{center}
\end{figure}
\begin{claim}[Duality]\label{cl:resistor-connection}
    Let $\flownet=(G, \vec{F}, \vec{I})$ be a conservative supply network with flows $F_{ij}=K_{ij}f(\varphi_j-\varphi_i)$ across each edge as per
    \eqref{eq:conserv-prop-flow}. Suppose the flow across the edge $(a,b)$ is positive from $a$ to $b$, $F_{ba}>0$, and let its strength be increased as per Definition 
    \ref{def:be}. Let the resulting flow changes
    across each edge be $F'_{ij}=F_{ij}+\delta F_{ij}$. Now consider a
    resistor network $\metaGraph{G}$ that has the same vertex and edge sets as $G$, and each edge $(i,j)\in\E$ has resistance
    \begin{equation}
        \label{eq:red-capacity}
        1/\tilde{K}_{ij}=1/K_{ij}f'(\varphi_j-\varphi_i).
    \end{equation}
    Suppose a constant dipole current source
    with current $I=\frac{F_{ab}\kappa}{K_{ab}}$ is connected across the edge $(a,b)$
    so that $b$ is the input node and $a$ is the output node (i.e. 
    anti-aligned to the original flow there).  Then
    the current $I_{ij}$ across any edge $(i,j) \neq (a,b)$ equals
    $\delta F_{ij}$ given by Definition  \ref{def:be}.
\end{claim}
Figure \ref{fig:resistor-equiv} illustrates that the incremental flow changes due to increasing the strength of an edge (the right edge in 
panel a) equals the currents induced by a single dipole current in the resistor network in panel b. The duality implies that the right edge must be Braessian, because at the left edge (maximum flow), the current (panel b) is aligned to the 
original flow (panel a). A detailed derivation of the duality is provided in Appendix \ref{appsec:resistances}.

The importance of this resistor network equivalence lies in a symmetry the
resistor problem possesses: the current at an edge $\mu$ due to a
constant current source (CCS) source at $\nu$ equals the current at 
$\nu$ due to another CCS at $\mu$ (For a review of this concept with  details, we refer to  Appendix \ref{appsec:resistor-symmetry}). This symmetry is  the discrete 
analog of symmetry present in systems of one dipole placed in a continuous electric field: the electrical field measured at $\vec{x}$ 
due to an electrostatic dipole placed at $y$ is identical to the electrical field at $\vec{y}$ due to a dipole placed at $\vec{x}$. This symmetry results in the following lemma. 

\begin{figure}[!htp]
    \begin{center}
        \includegraphics[width=\columnwidth]{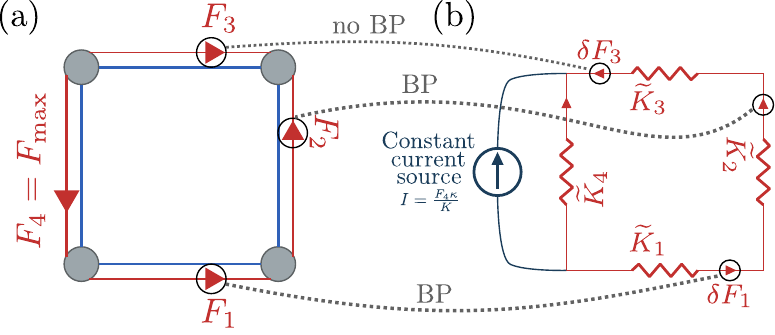}
        \caption{Detecting all Braessian edges in a network using the resistor
        equivalence. The edges for which the original flow and the current
        flow in the resistor network are \emph{in the same direction} are Braessian. The edges for which these two are in the
        opposite direction are not Braessian. }
        \label{fig:be-detect-smart}
    \end{center}
\end{figure}

\begin{lemma}\label{lem:resistor-symmetry}
    Consider a supply network $\flownet=(G,\vec{F},\vec{I})$ with maximum flow across edge $(s,t)$, directed from $s$ to $t$.  Consider a resistor network $\metaGraph{G}$ with a
    constant current source connected with the positive  terminal attached to
    $t$ and the negative terminal attached to $s$,  resulting in currents
    $I_{ab}$ for each edge $(a,b)\in G$. If $I_{ab}$ is directed identically
    as  the flow $F_{ab}$ in the original flow  network $\flownet$, i.e. $I_{ab}K_{ab}>0$, then, and only then, $(a,b)$ is a Braessian edge.
\end{lemma}
As a consequence, exploiting the symmetry of the resistor currents,
determining the Braessianness of \emph{all} the edges requires \emph{just 
one step}: place a dipole across the maximum flow, and compute the 
currents. The brute-force method would be to strengthen each edge \emph{one by
one} and compute the new steady flows. We illustrate this in
\fig\ref{fig:be-detect-smart}. 

We now demonstrate that the
resistor network analog, beyond a major speedup in numerical 
identification, enables us to gain an intuitive \emph{topological understanding} about which 
edges are likely to be Breassian.

\section{Topological understanding of Braess paradox}
\label{sec:alignment-idea}
If we had a graph theoretical quantity -- preferably easy to compute -- that
predicts the direction of current in a resistor network due to a constant
current source across one of its edges, our problem of predicting Braessian
edges based on topology would be completely solved, thanks to Lemma
\ref{lem:resistor-symmetry}.

As it happens, there exists such a quantity, presented by Shapiro
\cite{shapiro1987electrical}; which we will paraphrase here.

\begin{lemma}(Based on \citep[Lemma 1]{shapiro1987electrical})
    \label{lem:shapiro}
    Consider a resistor network with 1 unit of \footnote{Say, 1 Ampere} current across an edge $(s,t)$, 
    directed from $s$ to $t$. We are interested in finding out if the current 
    across an arbitrary edge $(a,b)$ is
    directed from $a$ to $b$ or from $b$ to $a$. Let $\N(s,a\to b,t)$ be the set of
    spanning trees containing a path $s,v_2,\cdots,a,b,\cdots,v_{m-1},t$.  Let
    $\N(s,b\to a,t)$ be defined in an analogous manner.   Then the current
    across $(a,b)$ is directed from $a$ to $b$ ($b$ to $a$) if
    \begin{align}
    \label{eq:shapiro-criterion}
        \sum_{T\in \N(s,a\to b,t)} \sum_{(i,j)\in T} R_{ij} \gtrless \sum_{T\in
        \N(s,b\to a,t)} \sum_{(i,j)\in T} R_{ij},
    \end{align}
    where $R_{ij}$ is the resistance of the edge $(i,j)$ and the sums run over the spanning trees $T$. 
\end{lemma}
Unfortunately the double sum in \eqref{eq:shapiro-criterion} is complex to
compute, hence not useful in our quest of predicting Braessian edges. We will thus
now present a simple topological concept we call \emph{rerouting
alignment}, inspired by \eqref{eq:shapiro-criterion}. It  is easy to compute, intuitive to understand, and frequently agrees with \eqref{eq:shapiro-criterion} to act as an approximate predictor of Braessian edges.

\begin{definition}[Rerouting alignment] \label{def:rerouting-alignment}
    Consider the same resistor network as in Lemma \ref{lem:shapiro}. Let
    $P_{s,a, b, t}$ be the shortest simple path that starts at $s$, ends at $t$
    and contains the edge $(a,b)$. If $a$ preceeds $b$ in
    $P_{s,a,b,t}$, then we say $(a,b)$ is aligned by rerouting to $(s,t)$.
    Otherwise, we say $(a,b)$ is anti-aligned by rerouting to $(s,t)$.
\end{definition}

Now state a Heuristic that in the setup described in Lemma 
\ref{lem:shapiro}, the current across the edge $(a,b)$ will be directed from 
$a$ to $b$ 
(from $b$ to $a$) if $(a,b)$ is aligned (anti-aligned) by rerouting to $(s,t)$.
Whenever this Heuristic holds, Lemma \ref{lem:resistor-symmetry}  yields
the following predictor for Braessian edges in a network:
\begin{conjecture2}
    Suppose the maximum flow is across the edge $(s,t)$, directed from $s$ to
    $t$.  Given any other edge $(a,b)$, carrying flow from $a$ to $b$, is
    Braessian if and only if it is aligned by rerouting to the edge $(s,t)$.
    \label{conj:conj}
\end{conjecture2}
\begin{figure}[!htp]
    \begin{center}
        \includegraphics[width=\columnwidth]{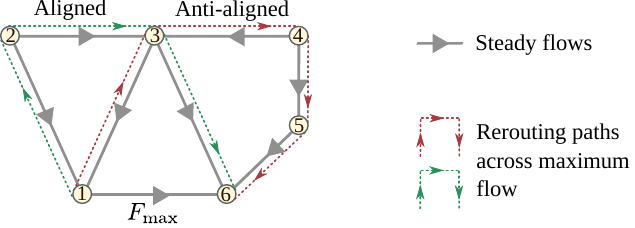}
        \caption{\textbf{Rerouting alignment as per Definition \ref{def:rerouting-alignment}}. Bottom edge $(1,6)$ is maximally loaded and the flow is from $1$ to $6$. Top left edge $(2,3)$ is aligned by rerouting to $(1,6)$: the shortest rerouting path from $1$ to $6$ containing $(2,3)$ is $(1,2,3,6)$, which aligns with the direction of flow across it: $2 \to 3$. But top right edge $(3,4)$ is anti-aligned by rerouting to $(1,6)$: the shortest rerouting path from $1$ to $6$ containing $(3, 4)$ is $(1,3,4,5,6)$, which does not align with the direction of flow across it, which is from $4 \to 3 $.}
        \label{fig:classifier-diagram}
    \end{center}
\end{figure}
\subsection{Accuracy of the topological predictor}
\begin{figure}[!htp]
    \begin{center}
        \includegraphics[width=\columnwidth]{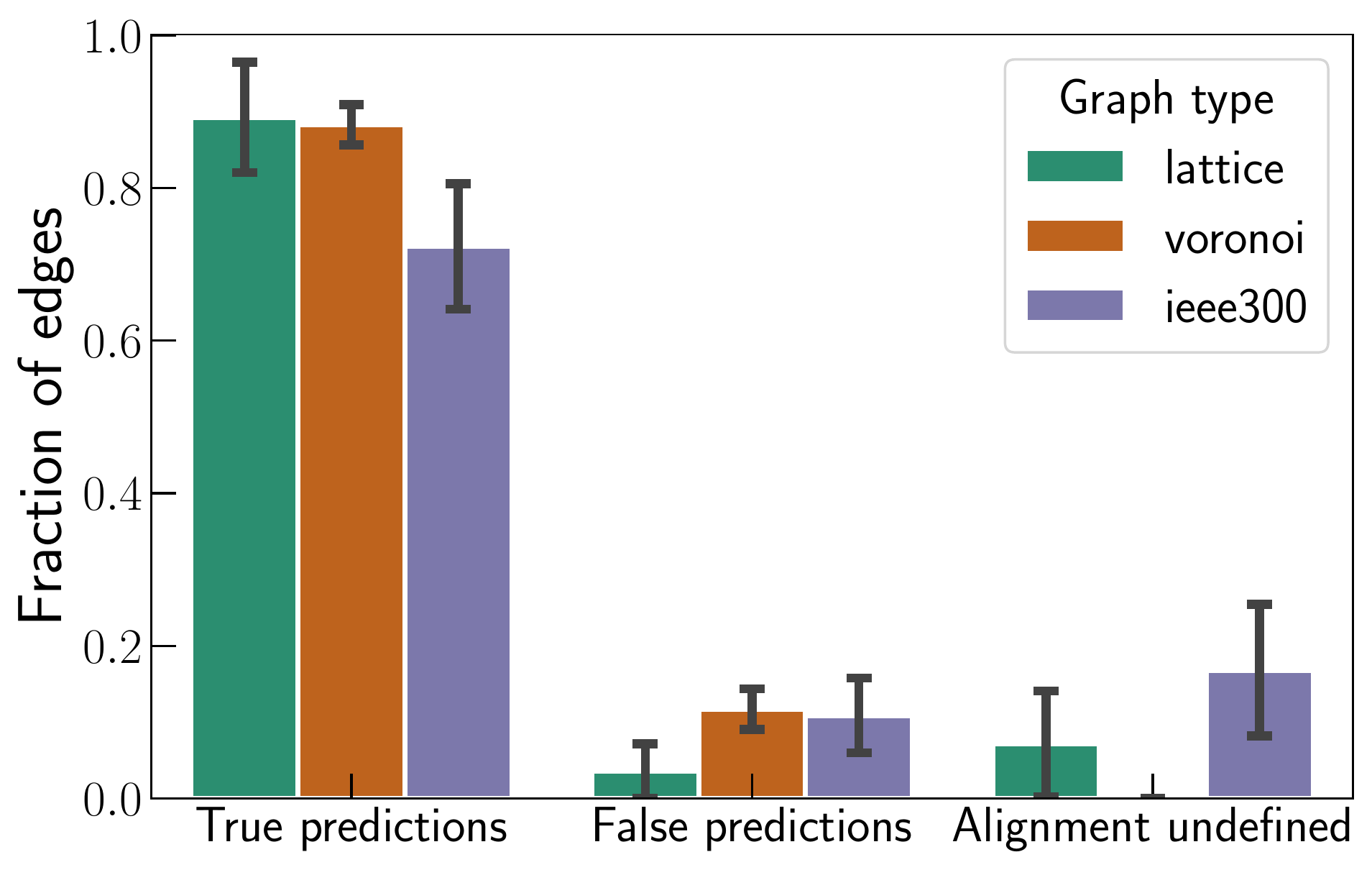}
        \caption{Braessian Edge classifier demonstrated to work in three network topologies:
        $15\times 15$ square lattice, Voronoi tessellation of $20$ uniformly
        randomly chosen points in the unit square, and the IEEE 300 bus test grid. 
        In each case $200$ independent input/output configurations were 
        examined. Error bars display the standard deviations. Edges that caused a change in maximum flow $\delta F_{st}\le 10^{-4}\kappa$, as well as the bridge edges were excluded from the analysis (see Appendix \ref{appsec:predictor} for details).}
        \label{fig:classif-two-networks}
    \end{center}
\end{figure}
Now, Heuristic \ref{conj:conj} does not always hold, but often, 
making it an effective predictor for Braessian edges.  To substantiate this claim, we analyzed its performance 
 in three classes of drastically different network topologies (\fig \ref{fig:classif-two-networks}): a
$15\times 15$ square lattice, a Voronoi tessellation of $20$ uniformly
randomly drawn points from a unit square and the IEEE 300 bus test case. In each topology, $1/4$th of the
nodes were chosen to have inputs $1$ and an equal number to have inputs $-1$.
The remaining nodes have inputs $0$. For all three topologies, we generated and analyzed $200$ independent
realizations. We find that the classifier based on the above Heuristic performs reasonably well. The exact implementation of the predictor is described in Appendix \ref{appsec:predictor}. 

\section{Heuristics for mitigating network overload}
\label{sec:mitigation}
Braessian edges, by their very definition, increase the maximum flow in the
network when strengthened. Vice versa, they \emph{decrease} the maximum flow
when weakened. Utilizing this property, we will now show how to \emph{mitigate}
overload in a network caused by damage at an edge by \emph{damaging a second,}
Braessian edge. We note that a similar phenomenon was reported in
\cite{motter2004cascade}, where intentionally removing certain nodes and edges
were shown to reduce the extent of cascading failures in a network.
\begin{figure}[!htp]
    \begin{center}
        \includegraphics[width=\columnwidth]{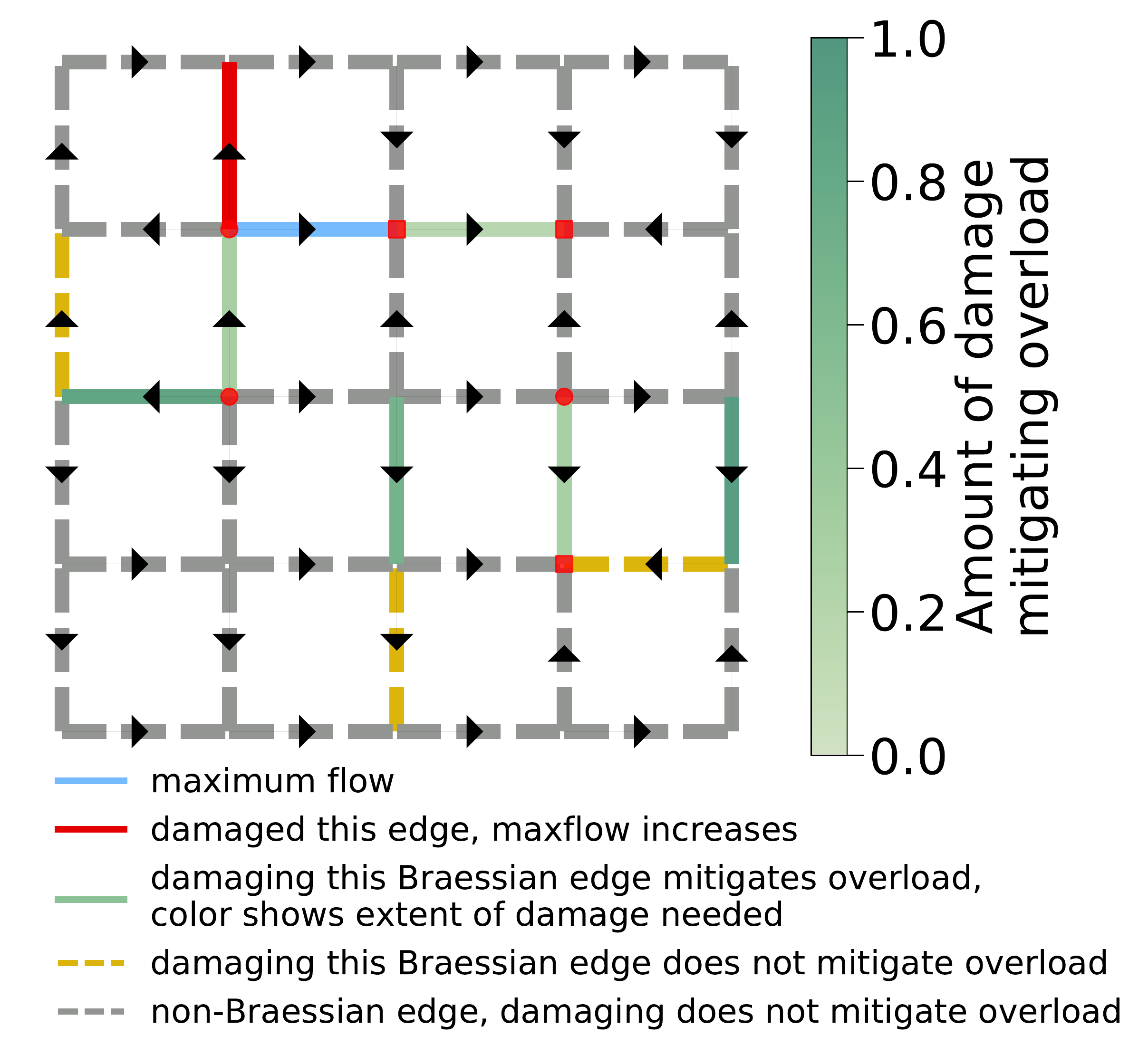}
        \caption{Braessian edges, when damaged, may mitigate overload in a
        network. The strength of the edge marked in red is reduced by 10\%,
        resulting in an increase of the maximum flow (across edge marked in
        blue). The edges in yellow, as well as shades of green are Braessian. Subsequently
        decreasing the strength of any of the green Braessian edges by suitable
        amounts reverts the increase of maximum flow. This approach cannot be
        used with the  Braessian edges marked in yellow. Non-Braessian edges
        are marked in gray.}
        \label{fig:be-mitigating}
    \end{center}
\end{figure} In Figure \ref{fig:be-mitigating}, we illustrate this for a
$5\times 5$ square lattice, with each edge having the same weight of unity,
$K_{ij}=1$. Reducing the strength of one edge (colored red) edge  by $0.1$
causes an overload in the maximum flow-carrying edge (colored sky blue). Among
the Braessian edges, many mitigated the overload, when damaged to a suitable
degree (by reducing their strengths). The colormap in the figure illustrates
the amount by which the weight of an edge must be reduced to bring the maximum
flow in the network back to its original value. Not coincidentally, the
non-Braessian edges were incapable of mitigating the overload by this strategy
of weight reduction.  However, some Braessian edges cannot mitigate the
overload, no matter how much they are managed.  According to our systematic
observations, this was due to one of two reasons. First, there were edges that,
even when damaged to the maximum degree (i.e. completely taken out), could not
completely reverse the overload. Secondly, there were some edges, which when
damaged suitably, although reversed the overload in the previously maximally
loaded edge, ended up overloading \emph{another edge} so much that the maximum
flow in the network increased.  The flow across another edge became the new
maximum flow.

\section{Conclusion}
In this article we have presented an intuitive and topological way of classifying which edges in a supply network exhibit Braess' paradox such that increasing their strength increases the maximum flow. In real world networks that often are capacity constrained, such increased maximum flows may easily induce overloads and system dysfunction.

Many supply networks crucial for our society need upgrading single edges from time to time. We thus believe that an improved intuitive understanding of the consequences of upgrading infrastructures may help planning such upgrades.  We have shown that the incremental flow changes upon an infinitesimal strength increase of an edge are
equivalent to the currents in a  suitably constructed resistor network. This equivalence may be exploited beyond predicting Braess' paradox, because  it contributes an intuitive understanding of how the flow across \emph{any} edge of choice would be affected if any other edge strength is changed. 

Moreover, the resistor network analog may be extended to understand the effect of changes at \emph{multiple} edges at once: It is equal to currents due to \emph{multiple dipole} current sources in the resistor network. The latter follows from the resulting linearity of the differential approach and therefore the superposition principle underlying the problem.

We have concentrated in this  article on infinitesimal increases in edge strengths, and it remains to investigate  how our results translate to settings with non-infinitesimal changes in edge  strengths, including a newly added or entirely removed edge, using, e.g., line outage distribution factors \cite{LODF2017dual}.

\textit{Acknowledgements.} We thank Rainer Kree, Franziska Wegner, Benjamin Sch\"afer, and Malte 
Schr\"oder for valuable discussions and hints about the manuscript presentation. We gratefully acknowledge support from
the Federal Ministry of Education and Research (BMBF grant no. 03EK3055A-F), the International Max Planck Research School for the Physics of 
Biological and Complex Systems (to DM),
the German Science Foundation (DFG) by a grant toward the Cluster of Excellence 'Center for Advancing Electronics Dresden' (cfaed).

\textit{Code availability.}
Code to reproduce key results is available at \cite{self_Code_to_Reproduce}.

\bibliography{Braess,networks}

\begin{appendix}

\section{Flows in a resistor network}
\label{appsec:resistances}
    We demonstrate here how \emph{incremental flow changes} upon 
    strengthening an edge in a conservative supply network are equivalent to 
    the electrical currents in a suitably constructed DC resistor network.
    Suppose a 
    resistor network is described by a graph $G(\V,\E)$, with each edge $(i,j)$ 
    having resistance $1/K_{ij}$.  Let the input of electrical current at each node $j$ be $P_j$. Then the input at each node $j$ must equal the total outwards current from $j$ to all its neighbours, i.e.  
\begin{equation}
    P_j = \sum_{(i,j)\in\E} F_{ij}, \text{ for all } j \in\V,
    \label{res-conserv}
\end{equation}
meaning the continuity equation \eqref{eq:flowcon} is satisfied. In addition, Ohm's law gives
\begin{equation}
    \label{res-edge}
    F_{ij} = K_{ij}(V_j -V_i),
\end{equation} 
where $V_j$ is the voltage at node $j$.

\subsection*{Flows in a resistor network due to a single constant current 
source}
Suppose in a resistor network, a constant current source with current $I_0$ is 
placed across edge $(a,b)$ so that $P_j=I_0(\delta_{jb}-\delta_{ja})$.  

Now combining \eqref{res-conserv} and \eqref{res-edge}, we can see that
\begin{equation}
    \label{origflows-resistor}
\begin{aligned}
    F_{ij} &= K_{ij}(V_j -V_i) \\
    P_j & = \sum_{(i,j)\in\E} K_{ij} \left(V_j - V_i\right) = I_0(\delta_{jb}-\delta_{ja}).
\end{aligned}
\end{equation}

\subsection*{Relation with flow changes on infinitesimal strengthening of an edge}
Now we  justify our claim \ref{cl:resistor-connection} that in any conservative
supply network, if 
the strength of an edge is infinitesimally increased as per Definition \ref{def:be}
, the resulting flow changes across any edge will be equal to currents in a suitably constructed 
resistor network.

In a supply network $\flownet=(G, \vec{I}, \vec{F})$, combining 
\eqref{eq:flowcon} and \eqref{eq:conserv-prop-flow}, we see that the flows 
will satisfy
\begin{align}
    \label{eq:edge-balance}
    I_j &= \sum_{(i,j)\in\E} F_{ij} = \sum_{(i,j)\in\E} K_{ij}f(\varphi_j - \varphi_i).
\end{align}

Now if the strength of a single edge $(a,b)$ is increased from $K_{ab}$ to 
$K_{ab}+\kappa$, let the $\varphi_j$ at each node $j$ be changed to 
$\varphi_j+\xi_j$. Then the new flows will be
\begin{equation}
    \label{new-flows}
    F_{ij} + \delta F_{ij} = \left(K_{ij}+ \kappa\delta_{ai}\delta_{bj} + \kappa\delta_{aj}\delta_{bi}\right) f\left(\varphi_j+\xi_j - \varphi_i - \xi_i\right).
\end{equation}
Defining 
\begin{equation}
    \label{def-ktilde}
    \tilde{K}_{ij}= K_{ij} f'(\varphi_j-\varphi_i),
\end{equation}
we see that the flow changes at all edges $(i,j)\neq(a,b)$ will be
\begin{equation}
    \label{flowchnages}
    \delta F_{ij} = \tilde{K}_{ij} \left(\xi_j - \xi_i\right).
\end{equation}

With the new flows \eqref{new-flows}, \eqref{eq:edge-balance} will become 
\begin{align*}
    I_j &= \sum_i \left(K_{ij} + \kappa\delta_{ai}\delta_{bj} + 
    \kappa\delta_{aj}\delta_{bi} \right) f(\varphi_i - \varphi_j + \xi_i - \xi_j)\\
       &= I_j + \kappa\delta_{bj}f(\varphi_a-\varphi_j) + 
      \kappa\delta_{aj} f(\varphi_b - \varphi_j)  +\sum_i 
      \left[K_{ij}\right.    \\ 
      &\left. \left\{f(\varphi_i - \varphi_j) +f'(\varphi_i - \varphi_j)(\xi_i -
       \xi_j)\right\}\right] + \bigO{(\xi_i-\xi_j)^2}.
\end{align*}
Subtracting Eq.~\eqref{eq:edge-balance}:
\begin{align*}
    0 &= \kappa\delta_{bj}f(\varphi_a-\varphi_j) + 
      \kappa\delta_{aj} f(\varphi_b - \varphi_j) + \sum_i \tilde{K}_{ij}(\xi_i - \xi_j) \\
      & + \bigO{(\xi_i-\xi_j)^2}.
\end{align*}

Rearranging and putting together with \eqref{flowchnages} we see
\begin{equation}
    \label{equiv-final}
\begin{aligned}
    \delta F_{ij} &= \tilde{K}_{ij} \left(\xi_j - \xi_i\right)\\
    \sum_i \tilde{K}_{ij}(\xi_j - \xi_i) &= \kappa f(\varphi_a-\varphi_b) 
    \left(\delta_{jb}-\delta_{ja}\right) \\
    &=\frac{F_{ab}\kappa}{K_{ab}}\left(\delta_{jb}-\delta_{ja}\right)
\end{aligned}
\end{equation}

Comparing the flow changes \eqref{equiv-final} and the currents in a resistor 
network \eqref{origflows-resistor} yields  Claim 
\ref{cl:resistor-connection}. That is, we see that the flow changes upon 
increasing the strength of edge $(a,b)$ with original flow $F_{ab}$ 
directed from $a$ to $b$; the resulting flow changes $\delta F_{ij}$ across 
all edges $(i,j)\neq (a,b)$ is given by the electrical currents in a resistor  network with the same topology, and each edge having resistance 
$1/\tilde{K}_{ij}$, and a single constant current source with current 
$I=\frac{F_{ab}\kappa}{K_{ab}}$ placed across the enhanced edge $(a,b)$, such 
that $b$ is the current source and $a$ is the sink. This equivalence is
illustrated in \fig \ref{fig:resistor-equiv} for a simple example.

\section{A symmetry of resistor currents}
\label{appsec:resistor-symmetry}

\begin{figure}[!htp]
    \begin{center}
        \includegraphics[width=\columnwidth]{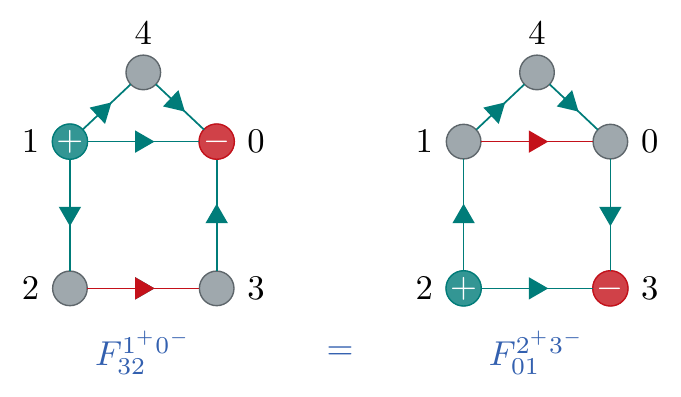}
        \caption{Symmetry of current in a DC resistor network.  Flow from 
        vertex $2$ to $3$ due to a constant current source placed across $(1,0)$ 
        (positive at $1$, negative at $0$) is 
        the same as the current from $1$ to $0$ due to a constant current 
        source placed across $(2,3)$ (positive at $2$, negative at $3$).  
        }  
        \label{fig:dipole-equivalence}
    \end{center}
\end{figure}

In a resistor network, let a constant current source (CCS) be placed across the 
edge $(a,b)$ ($+I$ input at $b$, $-I$ at $a$) and the resulting
current at edge $(i,j)$ from $i$ to $j$ be $F_{ji}^{b^+a^-}$.  We  
show that if we \emph{swapped} $(a,b)$ and $(i,j)$ simultaneously, i.e.  placed a constant 
current source across $(i,j)$ with a suitably chosen current,  the current 
across $(a,b)$ satisfies 

\begin{equation}
    \label{equi}
    F_{ji}^{b^+a^-}=F_{ab}^{i^+j^-}.
\end{equation}

To demonstrate this, we will go back to the definition equations for currents 
in a resistor network \eqref{origflows-resistor}
\begin{equation}
    \label{repeat:rsistorflow}
\begin{aligned}
    F_{ij} &= K_{ij}(V_j -V_i) \\
    P_j & = \sum_{(i,j)\in\E} K_{ij} \left(V_j - V_i\right) = I(\delta_{jb}-\delta_{ja}).
\end{aligned}
\end{equation}
It is beneficial to recast this requation in matrix form.  To this end, we 
will introduce two vectors in $\mathbb{R}^n$ ($n$ being the number of nodes in $G$)
\begin{align}
    \label{def:simplevecs}
    \vec{\mathbb{1}} &= (\underbrace{1,1,\cdots,1}_{n\text{ times}})\\
    \vec{e}^i &= (\underbrace{0,0,\cdots,0}_{i-1 \text{ 
    times}},1,\underbrace{0,0,\cdots,0}_{n-i \text{ times}}),
\end{align}
and a matrix $L\in \mathbb{R}^{n\times n}$, the weighted Laplacian matrix \cite[p.~286]{godsil2001} of the graph 
$G$, the edge weigths being the inverse resistances
\begin{align}
    L_{i,j} = \left\{
   \begin{array}{c l }
       \displaystyle \sum_{(i,k)\in\E}K_{ik} & \; \mbox{if  $i=j$,}\\
       -K_{ij} & \; \mbox{if $i\neq j$ and $(i,j)\in\E$,}\\
       0  & \; \mbox{otherwise}.
  \end{array} \right.
\end{align}

Then \eqref{repeat:rsistorflow} becomes
\begin{equation}
    \label{resistflow-matrix}
    \vec{P} = I\vec{e}^b-I\vec{e}^a = L \vec{V}.
\end{equation}
Now, \eqref{resistflow-matrix} does not have an unique solution for  $\vec{V}$
because $L$ is singular, with a null space of dimension $1$ spanned by the 
vector $\vec{\mathbb{1}}=\left(1,1,\cdots,1\right)$.  This is no
surprise, since the voltages in a DC resistor network are defined up to an
arbitrary additive constant. Following
\cite{james1978generalised,wiki:moorepenrose}, the node voltages are given by
\begin{align}
    \label{eq:voltage_vecsol}
    \vec{V} &= \linv\vec{P}+c\vec{\mathbb{1}},
\end{align}
where $\linv$ is the Moore-Penrose pseudoinverse of $L$ and $c$ is any real 
number. Then substituting \eqref{resistflow-matrix} into 
\eqref{eq:voltage_vecsol}, we obtain the voltage at any node $m$.
\begin{align*}
    \vec{V} &= I \linv (\vec{e}^b - \vec{e}^a) \\
    V_m &= I \sum_{k=1}^n\left(\linv_{mk} \vec{e}^b_k - \linv_{mk} \vec{e}^a_k 
    \right) +c\\
        &= I \left(\linv_{mb} - \linv_{ma}\right) + c.
\end{align*}

Then the current $F_{ji}^{b^+a^-}$ from $i$ to $j$, following 
\eqref{repeat:rsistorflow}, is

\begin{align*}
    F_{ji}^{b^+a^-} &= K_{ij}\left(V_i - V_j\right) \\
    &= K_{ij} I \left[\left(\linv_{ib} - \linv_{ia}\right) - \left(\linv_{jb} 
    - \linv_{ja}\right)\right] \\
    &= K_{ij} I \left(\linv_{ib} - \linv_{ia} - \linv_{jb} + \linv_{ja}\right).
\end{align*}

Now, analogously, if a constant current source with is placed across $(i,j)$ 
with $+I$ input at $i$ and $-I$ input at $j$, then the current from $b$ to $a$ 
will be
\begin{align*}
    F_{ab}^{i^+j^-} &= K_{ab} I \left(\linv_{bi} - \linv_{bj} -\linv_{ai}+\linv_{aj} \right).
\end{align*}

Since $\linv$ is a symmetric matrix (for proof, see \cite[Lemma 
6.A.1]{dmanik_phdthesis}), we have
\begin{equation}
    \label{proof-symm-current}
    F_{ab}^{i^+j^-} =\frac{K_{ab}}{K_{ij}}  F_{ji}^{b^+a^-}.
\end{equation}

The prefactor $\frac{K_{ab}}{K_{ij}}$ is a positive constant, and equals zero if the strength of the constant current source placed across $(i,j)$ 
is $I\frac{K_{ij}}{K_{ab}}$. 

More importantly, \eqref{proof-symm-current} states the following regarding the 
\emph{direction} of currents in resistor networks: If due to a constant 
current source across $(a,b)$ with positive input at $b$ and negative input at 
$a$, the resulting flow across edge $(i,j)$ is directed from $i$ to $j$, then a constant current source with positive input at $i$ and negative 
input at $j$ yields a current across $(a,b)$ directed from 
$b$ to $a$.  This is illustrated in \fig\ref{fig:dipole-equivalence}, and 
provides Lemma \ref{lem:resistor-symmetry}.

\section{Implementation of the predictor (heuristic for rerouting alignment)}
\label{appsec:predictor} The Heuristic \ref{conj:conj} introduced above not only helps understanding the origin of Braessian edges (and non-Braessian ones) but also enables us to develop an
algorithm for (approximately) predicting Braessian edges in any conservative supply network. The
crucial part of this algorithm is determining if in a supply network an edge
$(u,v)$ is aligned by rerouting to the maximum flow at edge $(s,t)$, directed
from $s$ to $t$.

We note that the concept of an edge $(u,v)$ being aligned by rerouting to
another edge $(s,t)$ is \emph{undefined} if either of these two edges is a
\emph{bridge}. We call an edge $(i,j)$ a bridge if an originally connected network becomes disconnected by removing that edge. Such edges by
definition are not part of any cycle, and therefore, do not support any
rerouting flow. Indeed, strengthening them has no impact on the flows across other edges. Since they are therefore not interesting for the present article,  such edges have been excluded from all analyses in this article. While generating random networks, we also have ignored realizations where the
maximum flow itself is across a bridge. 

Algorithm \ref{algo:predicttopo} describes the predictor in detail. However,
the step of determining the shortest path in a graph that traverses the nodes
$s,i,j,t$ in that order proved too computationally expensive to solve exactly.
Therefore we have developed a heuristic for determining approximations for such paths. As illustrated in \fig\ref{fig:classif-two-networks}, that heuristic works reasonably well. The implementation of this heuristic is
available in \cite{self_Code_to_Reproduce}.

\begin{algorithm}[H]
\SetAlgoLined
\KwData{$(G(\V, \E), \vec{I}, \vec{F}^{\textsf{con}})$, a conservative supply network.\\
The maximum flow $F_{\max}$ is across edge $(s,t)$, from $s$ to $t$.}
\KwResult{Set of predicted Braessian edges $E_{be}$,\\
set of predicted non-Braessian edges $E_{nbe}$.\\
}
\tcc{algorithm}
$E_{\textsf{be}}\gets\{\}$\;
$E_{\textsf{nbe}}\gets\{\}$\;
$H\gets G$\;
\For(\tcp*[f]{set edge weights of $H$}){$(i,j)\in \E$}{
    Weigth of $(i,j)=K'_{ij}\gets K_{ij}f'(\varphi_j-\varphi_i)$, as per \eqref{eq:red-capacity}\;
}

\For(\tcp*[f]{flow is from $i$ to $j$}){$(i,j) \in \E$}{
    \eIf{there exists any simple path in $H$ containing the nodes $s,i,j,t$ in that order}
    {$l_{\textsf{al}}\gets$ (heuristically) the length of the shortest such path;}
    {$l_{\textsf{al}}\gets\infty$}
    \eIf{there exists any simple path in $H$ containing the nodes $s,j,i,t$ in that order}
    {$l_{\textsf{nal}}\gets$ (heuristically) the length of the shortest such path;}
    {$l_{\textsf{nal}}\gets\infty$}

    \eIf{$l_{\textsf{al}} < l_{\textsf{nal}}$}
    {
       $E_{\textsf{be}}\gets E_{\textsf{be}}\cup \{(i,j)\}$\;
    }
    {
        \eIf{$l_{\textsf{al}} > l_{\textsf{nal}}$}
        {
            $E_{\textsf{nbe}}\gets E_{\textsf{be}}\cup \{(i,j)\}$\;
        }
        {Braessianness of the edge cannot be predicted.}
    }
}
\caption{Topological predictor for Braessian edges}
\label{algo:predicttopo}
\end{algorithm}

\end{appendix}
\end{document}